\documentstyle[11pt,newpasp,twoside,epsf,txfonts]{article}

\def\edcomment#1{\iffalse\marginpar{\raggedright\sl#1\/}\else\relax\fi}
\reversemarginpar

\begin{document}

\title{Extinction Properties of Some Complex Dust Grains}
 \author{Anja C.\,Andersen}
\affil{NORDITA, Blegdamsvej 17, DK-2100 Copenhagen, Denmark}
\affil{Department of Astronomy \& Space Physics, Uppsala University, P.O.Box 515, SE-751 20 Uppsala, Sweden}
\author{Juan A.\,Sotelo}
\affil{Dpto. de Fisica, Informatica y Matematicas, Universidad Peruana Cayetano Heredia, Aptdo. 4314, Lima, Peru}
\author{Gunnar A.\,Niklasson}
\affil{Department of Materials Science, Uppsala University, P.O.Box 534, SE-751\,21 Uppsala, Sweden}
\author{Vitaly N.\,Pustovit}
\affil{Department of Radiology, Washington University, 4525 Scott. Ave., St. Louis, MO 63110, USA}

\begin{abstract} Dust particles in space may appear as clusters of individual grains. The morphology of these 
clusters could be of a fractal or more compact nature. To investigate how the cluster 
morphology influences the calculated extinction of different clusters in the wavelength 
range $0.1 - 100~\mu$m, we have preformed extinction calculations of three-dimensional clusters
consisting of identical touching spherical particles arranged in three different
geometries: prefractal,
simple cubic and face-centered cubic.
In our calculations we find that the extinction coefficients of prefractal and compact
clusters are of the same order of magnitude.
For the calculations, we have performed an in-depth comparison of the
theoretical predictions of extinction coefficients of multi-sphere clusters
derived by rigorous solutions, on the one hand, and popular discrete-dipole
approximations, on the other. This comparison is essential if one is
to assess the degree of reliability of model calculations made with the
discrete-dipole approximations, which appear in the literature quite frequently
without an adequate accounting of their validity.  \end{abstract}

\section{Introduction}

The shape of interstellar and circumstellar grains is still an outstanding 
issue. The complexity of the electromagnetic scattering 
problem limits the theoretical modeling of the shapes which can be studied to
 spheres, infinite 
cylinders and spheroids. However, the shape of many interstellar
grains are expected to be non-spherical and maybe even highly irregular.
One way to deal theoretically with irregular particles and clusters of
dust grains is to assume that they consist of touching spheres.
With such an assumption it is possible to construct many distinctly different
morphologies which can then be compared with observations. 

The problem of evaluating the extinction efficiency ($Q_{\rm ext}$) is that of
solving Max\-well's equations with appropriate boundary conditions at the 
cluster surface. For a homogeneous single sphere a solution was formulated 
by Lorenz (1890) and Mie (1908)
and the complete formalism is therefore often 
referred to as the Lorenz-Mie theory. A complementary solution based on 
the expansion of scalar potentials was given by Debye (1909). A detailed
description of this exact electromagnetic solution can be found in
the book by Bohren \& Huffman (1983).  
For a review on exact theories and numerical techniques for computing
the scattered electromagnetic field by clusters of particles we
refer the reader to the textbook by Mishchenko et al.\ (2000).
For a comprehensive review on
the optics of cosmic dust see Voshchinnikov (2002) and Videen \& Kocifaj (2002).

To investigate clustering effects we have computed and analyzed the extinction 
of different polycrystalline graphitic and silicate clusters. We have chosen 
clusters 
ranging from small to large, and which are either sparse or compact, 
to evaluate how the extinction is influenced by the structure. 
We focus on clusters consisting of 4, 7, 8, 27, 32 and 49
touching polycrystalline spheres with a radius of 10 nm. 
The extinction of the clusters is
calculated using two rigorous methods -- GA (G\'erardy \& Ausloos 1982), 
and the generalized multi-particle Mie (GMM) solution (Xu 1995; 1997) -- 
and two 
discrete dipole approximation (DDA) methods -- MarCoDES 
(Markel 1998) and DDSCAT (Draine \& Flatau 2000) -- 
 to test how well these latter approximations
 perform when applied to clusters with different morphology.  
DDSCAT is as such an exact solution if
enough dipoles are used in the approximation of the target. 
It has been used in a wide range of scattering problems concerning clusters
of particles including the extinction of
interstellar dust grains (e.g.\ Bazell \& Dwek 1990; Wolff et al.\ 1994; 
Stognienko et al.\ 1995; Fogel \& Leung 1998; Vaidya et al.\ 2001).
The rigorous solutions are only exact if a high enough 
number of multi-poles is treated. 

\section{Structure of the clusters}

 \begin{figure}
\vspace*{-3.2 cm}
\plotfiddle{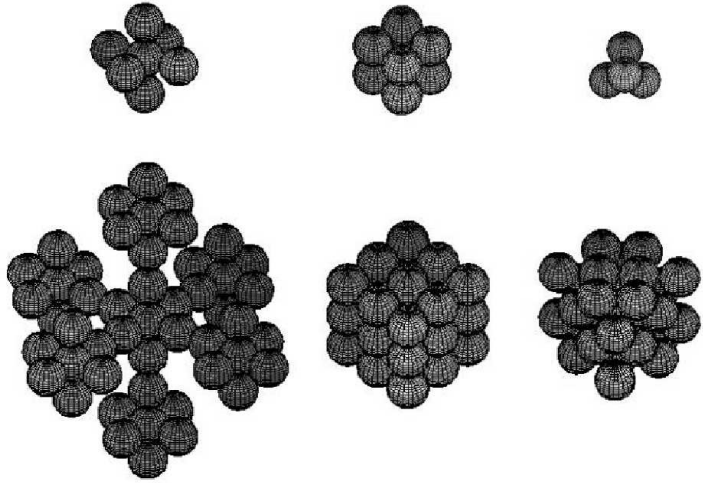}{10 cm}{-90}{40}{40}{-175}{210}
 \caption{Some of the clusters considered in the 
calculations.  From left to right: the prefractal clusters frac7 and frac49;
the simple cubic clusters sc8 and sc27; and the face-center-cubic clusters 
fcc4 and fcc32. The numerical suffixes indicate the number of spheres in the
clusters.} 
 \end{figure}

We consider three-dimensional clusters of identical touching spherical 
particles, of radius R, arranged in three different geometric configurations: 
prefractal (frac), simple cubic (sc), and face-centered cubic (fcc). 
These structures do not have shapes expected to be
found in space, but will provide us with boundary conditions
for the problem of calculating the extinction of clusters of grains
of different morphologies. 

The snowflake prefractal of $n$'th--order is recursively constructed
starting with the initiator which is just a single sphere. Next, the 1'st--order 
prefractal, or generator, is built up by pasting together seven copies of the
initiator as shown in the top left hand corner frame of Fig.\,1. For $n>1$, 
pasting together seven copies of the $n-1$'th order prefractal according to the
generator's pattern yields the $n$'th order prefractal. For example, the bottom
left hand corner frame of fig. 1 displays the snowflake prefractal of order 2.
As its order increases and goes to infinity, the snowflake prefractal
will become the snowflake fractal.

Vicsek (1983) has shown that regardless of their order all the snowflake 
prefractals have the same dimension, $D=\ln7/\ln3=1.77$, which is exactly the 
dimension of the snowflake fractal. This value is close to that obtained for
random cluster-cluster aggregation models for grain growth (Meakin 1988; 
Botet and Jullien 1988; Meakin \& Jullien 1988; Wurm \& Blum 1998). 

As a contrast to the prefractal structure, we also consider some compact 
crystalline structures, 
namely, face-centered cubic and simple cubic, see e.g. Kittel (1986) 
for a discussion on crystal 
structures. 
All of the clusters we consider are symmetric, 
see Fig.\,1, and consist of spheres with
radii $R=10$~nm.

\section{Material properties}

 \begin{figure}
 \plottwo{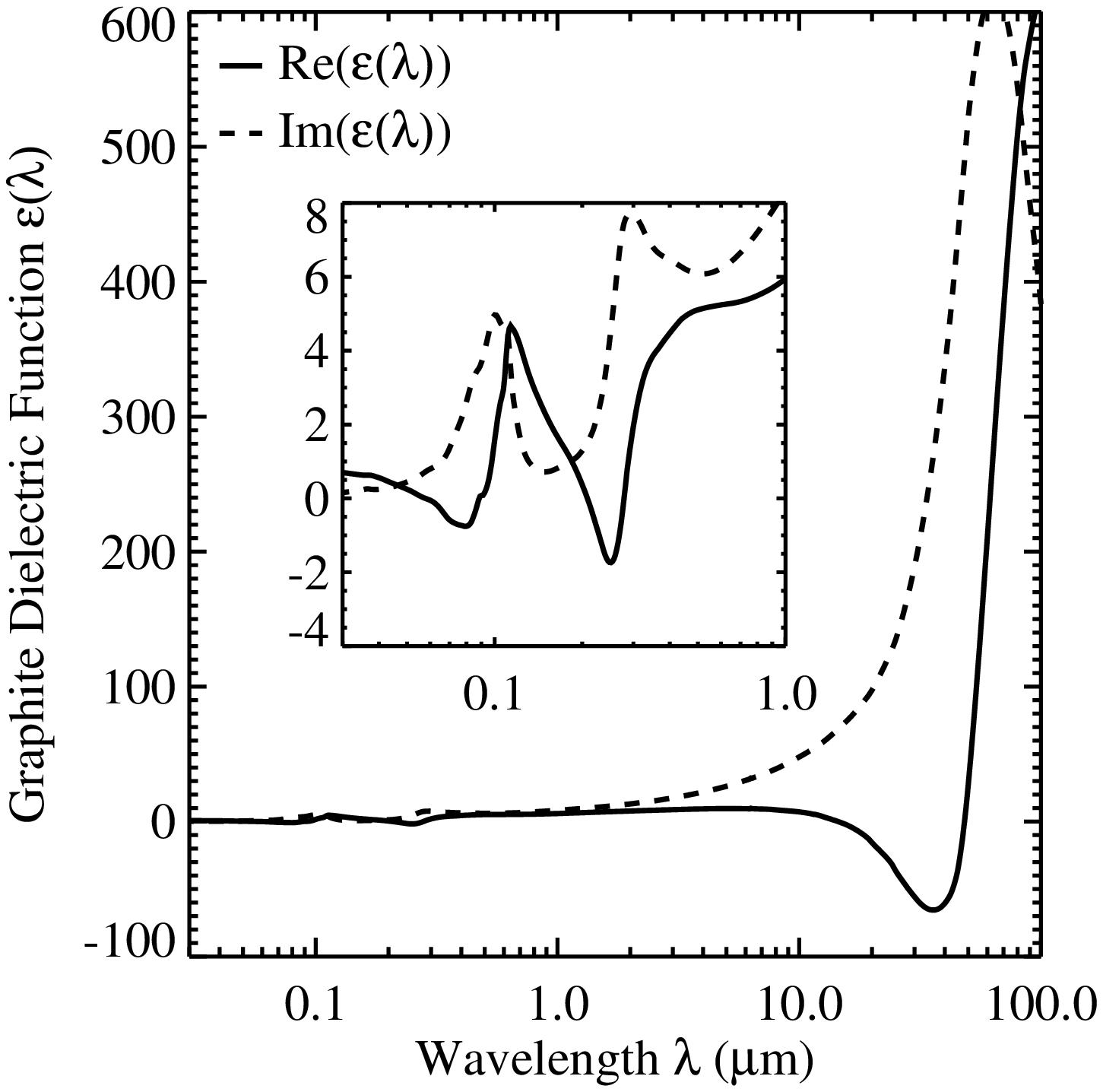}{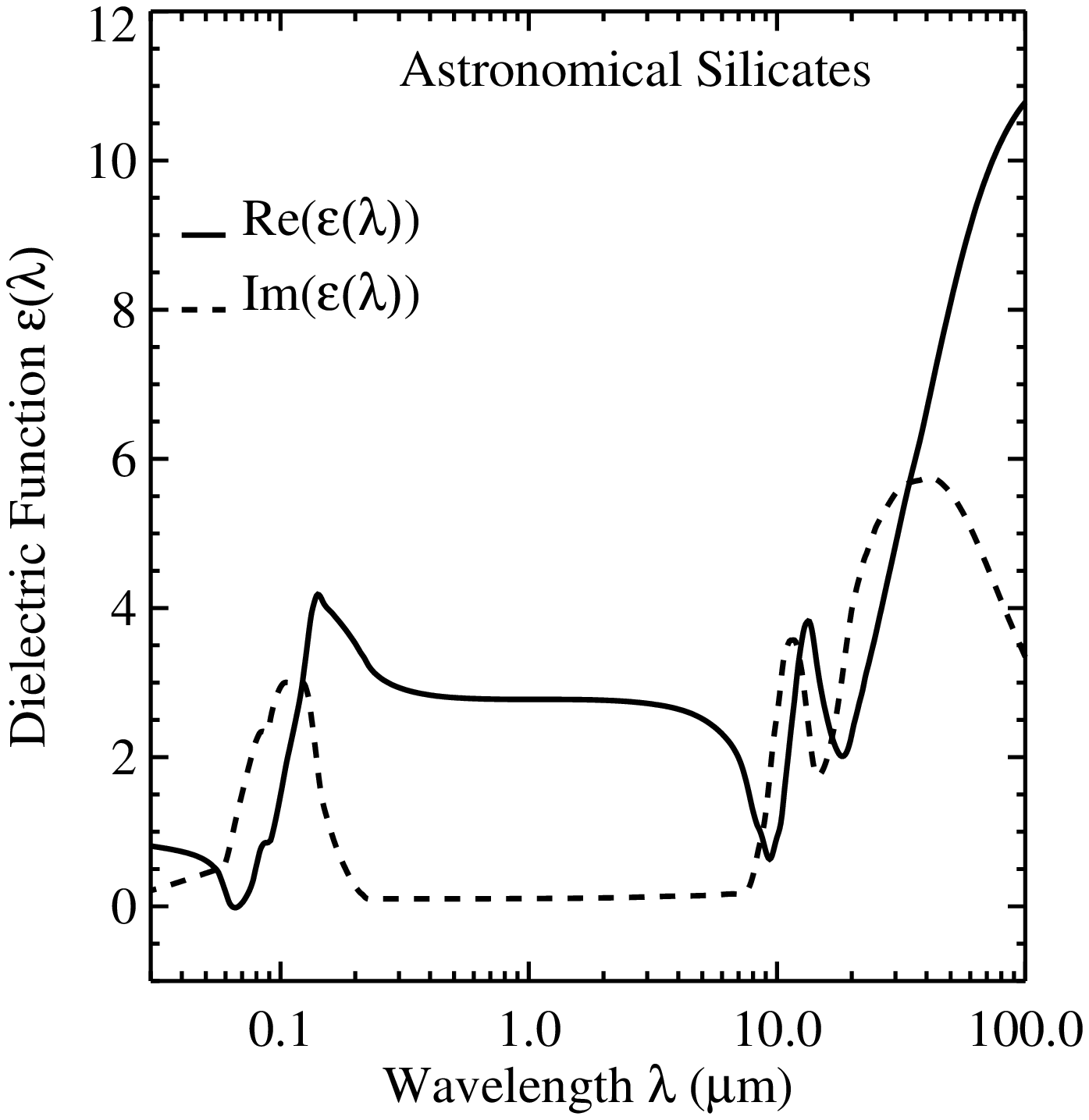}
 \caption{The dielectric functions, $\epsilon (\lambda)$, 
  of graphite (average) and astronomical silicates 
  used for the calculations of the extinction of the clusters.}
 \end{figure}

For our clusters we consider two different materials; graphite and silicates.
Graphite can be characterized by two different dielectric functions, 
$\epsilon_{\perp}$ and $\epsilon_{\parallel}$, corresponding to the
electric-field vector ${\bf E}$ being perpendicular ($\epsilon_{\perp}$) and
parallel ($\epsilon_{\parallel}$) to the symmetry axis of the 
crystal ($c$-axis), which is perpendicular to the basal plane. It is far
easier to experimentally determine $\epsilon_{\perp}$ than  
$\epsilon_{\parallel}$, 
because graphite cleaves readily along the basal plane and hence reflectivity
measurements can be made with normal incident light, in contrast,
it is very difficult to prepare suitable optical surfaces 
parallel to the $c$-axis. 
We use the dielectric functions $\epsilon_{\parallel}$ and
$\epsilon_{\perp}$ of graphite derived by Draine \& Lee (1984) covering 
the region from the far-IR to the far-UV.  
For the silicates we use the dielectric function of astronomical
silicates in the form given by Weingartner \& Draine (2001).
The dielectric functions of the two materials are shown in Fig.\,2.

As discussed by Draine \& Lee (1984) the dielectric constants for graphite are 
both temperature and size dependent. The graphite data obtained from Bruce 
Draine 
(http://www.astro.princeton.edu/$\sim$draine/dust/dust.diel.html) are given
for particle radius $R=100$\,nm. When using the data for other 
grain sizes it is necessary to correct the data according to Eq.\ 2 in 
Draine \& Lee (1984)\footnote{Notice that the plasma frequency for 
$\epsilon_{\perp}$ should be taken from the Errata (Draine \& Lee 1987).}.
The effect of the size corrections on the (average) optical constants
$n$ and $k$ are shown in Fig.\,3 for different
grain sizes. The correction is most significant in the long wavelength range
($\lambda > 50~\mu$m).

 \begin{figure}[t]
 \vspace*{-1.2 cm}
\plotfiddle{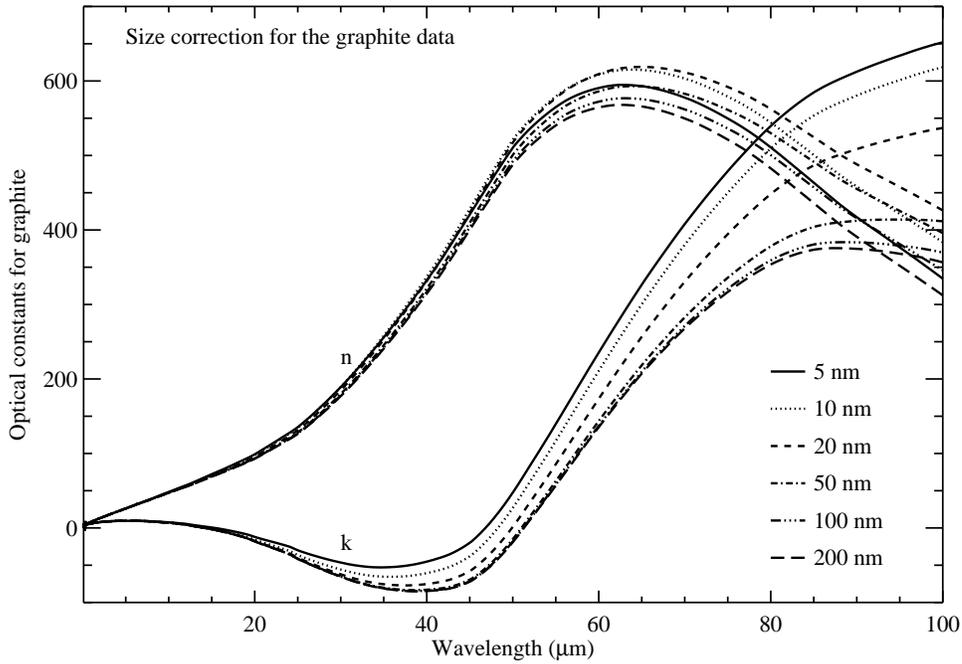}{10 cm}{0}{45}{45}{-230}{-35}
 \caption{The (average) optical constants $n$ and $k$ for graphite for 
different grain sizes.  The annotation within the figure refers to the 
radius, $R$, of the grain.}
 \end{figure}

In this work we deal with the anisotropy of graphite by assuming that in
all our clusters, each individual particle is polycrystalline having a
dielectric function $\epsilon_{\rm ave}$ given by the arithmetic average of
$\epsilon_{\parallel}$ and $\epsilon_{\perp}$, namely 
$\epsilon_{\rm ave} = \frac{1}{3} \epsilon_{\parallel} + \frac{2}{3} 
\epsilon_{\perp}$.
For a polycrystal the arithmetic average is a realistic model for $\epsilon_{\rm ave}$, since Avellaneda et al.\ (1988) have shown that 
it is an attainable upper bound for its dielectric function.  In stellar environments grains are most likely to grow as polycrystalline 
or even amorphous particles rather than as mono-crystalline (Gail \& Sedlmayr 1984; Sedlmayr 1994).  
In the usual ``1/3$-$2/3'' approximation each  individual particle is treated as a
mono-crystalline, where 1/3 of the cluster particles are assumed to have
dielectric function $\epsilon_{\parallel}$ and the remaining 2/3
to have dielectric function $\epsilon_{\perp}$. This approximation has been
shown by Draine (1988) and Draine \& Malhotra (1993)
to have a surprisingly good accuracy for graphite grains with
radii $\leq 200$~{\AA}.  However, for larger particles assuming
polycrystalline particles seems much more viable,  see e.g.\ Rouleau et al.\ (1997) 
for a discussion about different ways of obtaining an average dielectric function for graphite.

\section{The computational methods}

\subsection{The rigorous solutions}

Multi-particle scattering shows effects both of interaction and 
interference of scattered waves from the particles 
which can give rise to distinct features not seen in 
single-particle scattering. 

The two rigorous solutions presented here are generalized Mie theories
giving each a complete solution to
the multi-sphere light scattering problem. They are both based on the 
exact solution of Maxwell's equations for arbitrary cluster geometries, 
polarization and incidence direction of the exciting light.

\subsubsection{GA:} \label{ga}

A rigorous and complete solution to the multi-sphere light scattering 
problem has been given by G\'erardy \& Ausloos (GA) (1980;
1982; 1983; 1984) as
an extension of the Mie-Ruppin theory (Mie 1908; Ruppin 1975). 
The solution is obtained by expanding the various
fields involved in terms of vector spherical harmonics (VSH).
Boundary conditions are extended to account for the possible
existence of longitudinal plasmons in the spheres. High-order 
multi-polar electric and magnetic interaction effects are included. 

We consider a cluster of $N$ homogeneous spheres of radius $R$ and
dielectric function $\epsilon$, embedded in a matrix of dielectric
constant $\epsilon_M$ and submitted to a plane polarized time harmonic
electromagnetic field. The total scattered field from the cluster is
represented as a superposition of individual fields scattered from each
sphere. The electromagnetic field impinging on each sphere
consist of the external incident wave and the waves scattered by the other
spheres. For any sphere, the incident, internal and scattered fields are
expressed in VSH centered at the sphere origin.
The boundary conditions on its surface are
solved by transforming all relevant field expansions into the sphere
coordinate system, yielding a system of $2NL(L+2)$ equations whose solution
is the $2^L$-polar approximation to the electromagnetic response of the 
cluster. For a short description of this method see Andersen et al.\ (2002).

\subsubsection{GMM:} \label{gmm}

A neat analytical far-field solution to the electromagnetic scattering by an 
aggregate of spheres in a fixed orientation is provided by 
Xu (1995; 1997; 1998a; 1998b and Xu et al.\ 1999), and is 
implemented in the FORTRAN code gmm01f.f (GMM) 
available at http://www.astro.ufl.edu/$\sim$xu.
As any other rigorous solution to the multi-particle scattering, his 
approach considers two cooperative scattering effects: interaction and 
interference of scattered waves from individual 
particles (Xu \& Khlebtsov 2003). 
Nevertheless, his treatment of the second effect is novel. When a plane wave 
is incident upon the particles (scatterers) of a cluster, it has a phase 
difference determined by the geometrical  
configuration and spatial orientation of the cluster. Likewise, far  
away from the scatterers, the waves scattered from them also have well 
defined phase relations that depend on the scattering
direction. These incident and scattered phase differences give rise to 
far-field interference effects. Xu (1997) includes the incident-wave
phase terms in the incident-field expansions centered on each scatterer, and 
the scattered-wave phase terms in the 
single-field representation 
of the total scattered far-field from the whole cluster. This way of 
treating the interference effects is in practical calculations quite efficient
because then the required multipole order of the field expansions will depend 
only on the size 
of the individual particles and not on the
distance between them, that is, it will not depend on the 
size of the cluster (Xu \& Khlebtsov 2003). This allows in principle the 
treatment of clusters of arbitrary size; the
only limiting factor being the availability of computer memory. In general, 
an adequate estimate for the field-expansion 
truncation of all the 
scatterers in a cluster is given by the Wiscombe's criterion (Wiscombe 1980) 
for the field-expansion truncation of a single sphere
with size parameter $x$, $ L\approx x+4 x^{1/3}+2$. There 
are a number of cases, however, where this criterion 
grossly  underestimates the number
of multipoles needed in the scattering calculations. This is the case
for our clusters of graphitic spheres as it is for
the gold nano-bispheres discussed by Xu \& Khlebtsov (2003), and the 
soot bispheres discussed by Mackowski (1994). For example, to get a converged 
solution to the multi-particle scattering problem at wavelength 1.047 
microns for a cluster of 
8 graphitic spheres of radii 10\,nm, arranged in a simple cubic structure,
the actual single-sphere expansion truncation is 44 whereas the Wiscombe's 
criterion estimate is just 3. Finally, GMM implements 
two methods for solving the system of linear equations arising in multi-particle 
scattering, namely the order of scattering method of 
Fuller and Kattawar (1988a; 1988b) and the biconjugate gradient method (Gutknecht 1993).

\subsection{The discrete dipole approximations}

The discrete dipole approximation (DDA) - also known as 
the coupled dipole approximation - method is one of several discretization 
methods (e.g.\ Draine 1988; Hage \& Greenberg 1990) for solving scattering 
problems in the presence of a target with arbitrary geometry.
The discretization of the integral form of Maxwell's equations is usually 
done by the method of moments (Harrington 1968).  Purcell \& Pennypacker 
(1973) were the first to 
apply this method to astrophysical problems; since then, the DDA method has 
been improved greatly by Draine (1988), Goodman et al.\ (1991), 
Draine \& Goodman (1993), Draine \& Flatau (2000), Markel (1998), and 
Draine (2000). The DDA method has gained 
popularity among scientists due to its clarity in physical principle and the  
FORTRAN implementation which have been made publicly available by e.g.\
Draine \& Flatau (2000; DDSCAT) and by Markel (Markel 1998; MarCoDES). 

Within the framework of the DDA method, when considering the problem 
of scattering and absorption of linearly polarized light
\begin{equation}
E_{0} = e_{0} \exp \left( ikr \right) \label{vitaly1}
\end{equation}
of wavelength $\lambda=2\pi/k$ by an isotropic grain, the grain
is replaced by a set of discrete elements of volume $V_i$ with relative
dielectric constant $\epsilon_i$ and dipole moments $d_i=d\left(
r_i\right)$, $i=1,...,N,$ whose coordinates are specified by vectors $r_i$.
The equations for the dipole moments can be written
using simple considerations based on the concept of the exciting field,
which is equal to the sum of the incident wave and the fields of the rest of 
the dipoles in a given point. 

\subsubsection{DDSCAT:} 

In this work we use the Discrete Dipole Approximation Code version 5a10 
(DDSCAT; Draine \& Flatau 1994; Draine \& Flatau 2000), available at \\
http://www.astro.princeton.edu/$\sim$draine/DDSCAT.html. 
This version contains a new shape option where a target can be defined as the 
union of the volumes of an arbitrary number of spheres. 
In DDSCAT the considered grain/cluster is replaced by a cubic array
of point dipoles. The cubic array has numerical advantages because the
conjugate gradient method can be efficiently applied to solve the matrix
equation describing the dipole interactions (Goodman et al.\ 1991).
 
There are three criteria for the validity of DDSCAT: \\ \indent
(1) The wave phase shift $\rho =|m|ka$ over the distance $a$
between neighboring dipoles should be less than 1 for calculations of total
cross sections and less than 0.5 for phase function calculations.
Here, $m = \sqrt{\epsilon} = n+ik$ is the
complex refractive index of the target material \\ \indent
(2) $a$ must be small enough to describe the object shape 
satisfactorily. \\ \indent
(3) The refractive index $m$ must fulfill $|m| < 2$.

For materials with large refractive indexes ($|m| > 2$), Draine \& Goodman
(1993) have shown that especially the absorption is overestimated 
by DDA. As illustrated in Fig.\,2, graphite has a high refractive 
index throughout most of the range $0.03-100~\mu$m, showing that for graphite
the region of applicability of DDSCAT is rather small, in fact, 
the criterion $|m|<2$ is only
fulfilled for wavelengths shorter than $\lambda = 0.072~\mu$m ($m=0.58+i1.42$);
see the inset in the left frame of Fig.\,2. 
However, relaxing the criterion a little, to account for the variability 
of $m$ in the region below $1.0~\mu$m, the upper limit can be pushed up to
$\lambda = 0.216~\mu$m ($m=0.66+i1.35$). 

\subsubsection{MarCoDES:} 

Another efficient code based on DDA is the Markel Coupled
Dipole Equation Solver (MarCoDES; Markel 1998)
available at \\ http://atol.ucsd.edu/$\sim$pflatau/scatlib/. This code is
designed to approximate the spherical particles in an arbitrary cluster
with $point$ dipoles (this corresponds to $N=1$ in DDSCAT).  
The program is in principle applicable
to clusters of arbitrary geometry consisting of small spherical particles, 
but it is most efficient computationally for sparse clusters 
(i.e.\ when the volume fraction is very low) with significant number of 
particles $(\approx 10^3-10^4)$. Unlike DDSCAT, the program does not 
use the Fast Fourier Transformation (FFT) because this might significantly 
decrease its computational performance on
clusters with a low volume filling fraction. When the volume filling fraction
is close to unity, algorithms utilizing FFT will be much faster. 

The dimensionality of the coordinates of particles in
MarCoDES require a special consideration. By replacing real particles by
point dipoles located at their centers the strength of their interaction is
significantly underestimated. In order to correct the interaction strength,
the author of MarCoDES introduces geometrical intersection of particles.
All coordinates are defined  in terms of the distance between neighboring 
dipoles $a$, which is given by $a = \left( 4\pi /3\right)^{1/3} R$. 
So, for example, if two particles have radii 10~nm, then the
distance between the dipoles is 16.12~nm. This suggested
phenomenological procedure allows MarCoDES to be more accurate than the 
usual single dipole approximation since the intersection produces some
analogy of including higher multi-pole interactions between particles.
The fact that the program only uses a single dipole for each particle in the
cluster has significant benefits in computation efficiency when compared to
other multi-polar approaches such as GA, GMM or DDSCAT.

The version of MarCoDES tested here cannot calculate a 
face-centered cubic structure of touching particles because 
in this case the lattice cells representing neighboring particles will touch
only at the corners, giving as a consequence the spectrum corresponding to
non-touching particles.

\subsection{Accuracy of the different methods} 

Within the
GA method the extinction of a cluster is calculated in the $2^{L}$-polar 
approximation. In general, the smallest L needed for the convergence of the 
extinction differs for different regions of the optical spectrum; 
for graphite in particular, for a 
chosen fixed accuracy, the longer the wavelength, the  
higher the polar orders that are required in the calculations.
This happens because the magnitude of the refractive index 
of graphite increases with  
wavelength up to about $80.0~\mu$m, where it reaches a plateau. Generally, 
the extinction of graphitic clusters needs to be calculated to a 
higher polar order than that of the silicate
clusters to ensure convergence. 
In the UV-visible range, by accepting an accuracy 
of 5\% in the computation of the extinction, we can use L = 5 for 
open graphitic clusters and L = 7 for compact graphitic ones; we expect 
this to hold for clusters of up to a few tens of particles.  
In Table\,1, the cut-off polar order L used in the calculation of 
the extinction of all the clusters can be found. Full convergence was only 
achieved for the small clusters in the UV-vis range; for the larger 
clusters, L indicates the maximum polar order obtained with our 
available computer resources. 

\begin{table}
\caption{The clusters presented in this paper have three
different geometries (see Fig.\,1): prefractal (frac; D$=1.77$), face-center cubic 
(fcc) and simple cubic (sc). L designates the polar order achieved with the
GA method.}
\begin{center}
\begin{tabular}{|l||c|c|c|c|c|c|c|c|c|} \hline
Cluster Structure & frac & frac & fcc& fcc& fcc& sc& sc \\  \hline
\# particles & 7& 49& 4& 32& 49& 8& 27 \\  \hline
polar order L & 11 & 6 & 11 & 6 & 6 & 11 & 7 \\  \hline
Designation & frac7& frac49& fcc4& fcc32& fcc49& sc8& sc27 \\ \hline
\end{tabular}
\end{center}
\end{table}

 \begin{figure}
 \vspace*{-0.5 cm}
 \plotone{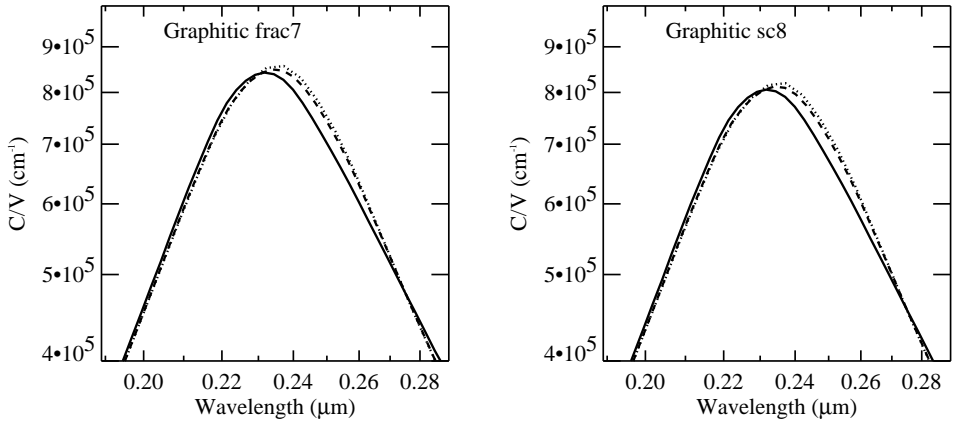}
 \caption{A comparison of the exact extinction calculation by GMM 
  (solid lines) for 
  the graphitic clusters frac7 and sc8 with the corresponding calculations 
  by DDSCAT using two different numbers of dipoles.  
  For frac7,
  the dotted line represents a calculation with
   $36\times36\times36$ = 46656 dipoles (giving 895 dipoles per particle) 
   while the dashed line represents a calculation with
   $48\times48\times48$ = 110592 dipoles (2103 dipoles per particle). 
    For sc8, the dotted line represents a calculation with 
   $32\times32\times32$ = 32768 dipoles (2103 dipoles per particle) 
   and the dashed line represents a calculation with 
   $36\times36\times36$ = 46656 dipoles (2969 dipoles per particle). 
   An increased number of dipoles leads, as expected, to an improvement 
   in the solution obtained with DDSCAT.}
 \end{figure}

The theory behind GMM is in many ways similar to that behind GA, 
differing from it in its use of an asymptotic form of the 
vector translational addition theorem in the
calculation of the total scattered wave in the far field, which 
avoids the severe
numerical problems encountered by GA when computing the latter 
for clusters of 
a large number of particles. As said earlier, GMM uses either the
order of scattering method or the biconjugate gradient method in solving 
the multi-particle scattering problem. Furthermore, when the former method 
fails in finding a solution, GMM switches to the latter method,
thus providing an answer in the majority of cases, although it may have 
to use very high polar orders to achieve a desired accuracy. For example,
to achieve an accuracy of four significant figures in the extinction
of the two small clusters frac7 and sc8, GMM needs to use polar 
orders as high as L=44 for wavelengths around $1.0~\mu$m. GA, on the 
other hand, proceeds one polar order at a time, and when it does 
not converge, it is not possible to establish the accuracy of the solution. 

 \begin{figure}[t]
 \vspace*{-1.0 cm}
 \plotfiddle{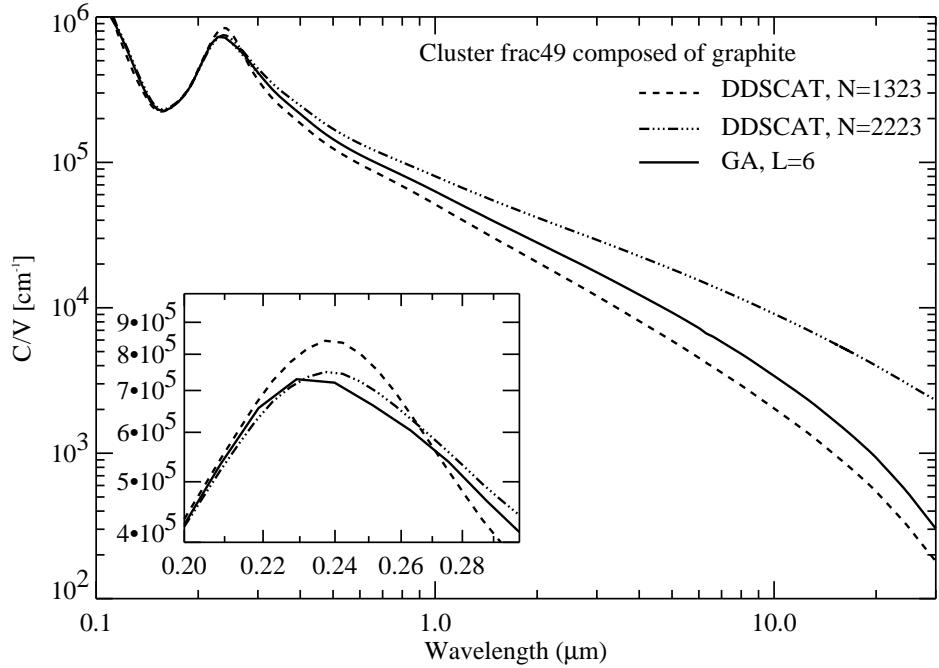}{10 cm}{0}{45}{45}{-220}{-30}
 \caption{Increasing the number of dipoles in a DDSCAT calculation will not
necessarily render a solution coming closer to the exact one. The plot 
shows the extinction of the graphitic cluster frac49 calculated, first, 
(dashed line) using 
$36 \times 36 \times 36$ = 46656 dipoles -- of which 1323
represented the target -- and then, (dot-dashed line) using
$40 \times 40 \times 40$ = 64000 dipoles -- of which 2223
represented the target. Also shown in the plot is the exact 
solution (solid line) as computed using GA with polar order L=6.} 
 \end{figure}

Xing \& Hanner (1997) find that the typical number of dipoles needed with 
DDSCAT to obtain a ''reliable'' computational result, 
can be determined by calculating the minimum number of dipoles needed per 
particle. When a particle of radius $R$ is represented by a
3-dimensional array of $N$ dipoles, its volume is $Na^{3}$, which
must be equal to $4\pi R^{3} / 3$, hence
\begin{equation}
N = \frac{4 \pi}{3} \left( \frac{R}{a} \right)^{3}  
  = \frac{4 \pi}{3} \left( \frac{2 \pi R |m|}{\rho \lambda} \right)^{3} 
  \approx 1039 \left( \frac{R |m|}{\rho \lambda} \right)^{3}
\end{equation}
since $a$ is related to the wave phase shift $\rho$ by $a = \rho / (|m|k)$
(Draine \& Flatau 2000).
For instance, for 
$\lambda \la 0.2~\mu$m, around 30 dipoles are needed for each of our 
graphitic spheres and for $\lambda \ga 0.2~\mu$m just one dipole 
seem to be enough, indicating that MarCoDES should be comparable 
with DDSCAT for wavelengths $\lambda \ga 0.2~\mu$m. 

As illustrated in Fig.\,4, for graphitic clusters increasing the
number of dipoles used in the DDSCAT calculation does lead to a solution
which is slightly closer to the exact result. However, for the frac7
cluster doubling the number of dipoles\footnote{The two calculations
used $36\times36\times36$ = 46656 dipoles of which 6265 dipoles 
represented the cluster (895 dipoles per particle) and  
$48\times48\times48$ = 110592 dipoles of which 14721 dipoles 
represented the cluster (2103 per particle).} only leads to a very slight
improvement of the solution. For example, the solution using the lower number of dipoles
is about 5\% off the exact solution around $\lambda = 0.24~\mu$m. 
Doubling the number of dipoles doubles the
computation time while the solution is only improved
by less than 1\%. According to Eq.\,(2) we are using more than an
adequate number of dipoles for both solutions indicating that a ''reliable''
result in the Xing \& Hanner (1987) terminology is less accurate than 5\%.

As seen from Fig.\,5 increasing the number of dipoles used in DDSCAT 
is not always a guarantee for getting a result which is closer to the exact 
solution. In the figure two different calculations of the frac49 cluster
composed of graphitic spheres are shown. The number of 
dipoles\footnote{The two calculations used $40 \times 40 \times 40$ = 
64000 dipoles and $36 \times 36 \times 36$ = 46656 dipoles. This resulted in
2223 dipoles and 1323 dipoles in the target and $\sim 45$ and 
$\sim 27$ dipoles per particle, respectively.} is almost doubled between
the two calculations. A comparison with the GA
calculations taken to polar order L=6, shows that almost doubling the number of dipoles
improves the result of the DDSCAT calculations by 10 \% around $\lambda = 0.24~\mu$m.
The GA solution is exact in the short wavelength range ($< 0.3~\mu$m) and coincides completely
with a calculation of GMM taken to L = 19. At longer wavelengths the GA
solution is not fully converged but we expect it to be in the vicinity
of the exact solution. The DDSCAT calculations show peculiar behavior at longer 
wavelengths since the solution using the lower number of dipoles comes
much closer to the GA solution than the solution using 30\% more dipoles.
This indicates that the lower number
of dipoles gives a better accuracy for longer wavelengths ($\lambda > 0.3~\mu$m) 
while the higher number of dipoles gives a better accuracy for shorter 
wavelengths.  In the case of
a discrete dipole array, the dipoles in the interior will be 
effectively shielded, while
the dipoles located on the target surface are not fully shielded and, as a 
result, absorb energy from the external field at an excessive rate. In 
principle the excess absorption which is introduced by having a large 
fraction of the dipoles at the surface of the particle should
be minimized when introducing more dipoles, but Fig.\,5 indicates that 
it is not necessarily a linear effect. It should be emphasized, however, that
the refractive index of the graphitic material have $|m| > 2$ at 
$\lambda > 0.3~\mu$m (see Fig.\,2) which means that the DDSCAT calculations 
are outside the range recommended by Draine \& Goodman (1993) for its use, 
so an overestimate of the absorption should be expected. Despite this,
increasing the number of dipoles should still lead to an improved
solution which is in contrast to what we find. This leaves open
the question of "how many dipoles are enough to assure a certain
accuracy in a DDSCAT computation". We therefore strongly recommend 
that this question gets investigated in much more detail in the future.

\section{Results}

 \begin{figure}
 \plotone{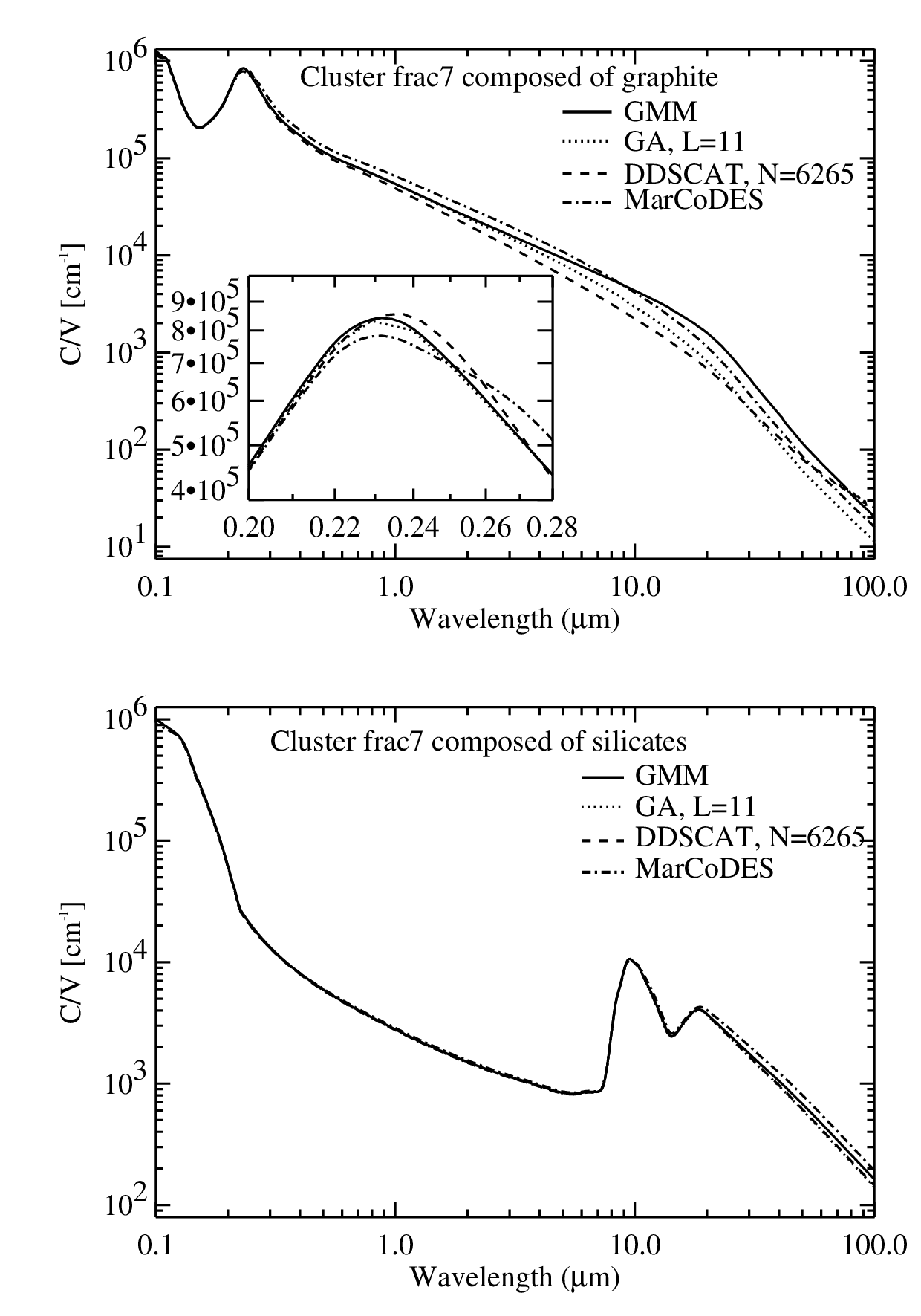}
 \caption{Cluster frac7 composed of graphitic material and silicates.
 The GMM solution has been computed with up to L=45 at certain 
 wavelengths; the GA, with up to L=11. The DDSCAT calculation was done with 
$36\times36\times36 = 46656$ dipoles (equivalent to 6265 dipoles within the 
target and $\sim 977$ dipoles per particle). }
 \end{figure}

To set up a comparison baseline, we computed the extinction of single
graphitic and silicate 
spheres, both of radius 10 nm, using the two
rigorous solutions (GA and GMM) and the 
two DDA codes (DDSCAT and MarCoDES). Since for a single sphere both the GA 
and GMM theories reduce to the Mie theory, the GA results are exactly
the same as those of GMM and equal to the Mie solution, regardless 
of the sphere's material. The two DDA 
codes, however, give results that differ markedly for both graphite  and
silicate.  For graphite, DDSCAT and the Mie solution coincide
up to $\lambda \la 1.0~\mu$m, where from DDSCAT starts to diverge slowly 
from the Mie solution. In contrast, MarCoDES differs from the 
Mie solution in the region $\lambda \la 0.6~\mu$m. This indicates that
for graphite DDSCAT might be the better choice of code for wavelengths 
$\lambda \la 1.0~\mu$m while MarCoDES is the better choice of code for 
longer wavelengths. 
For the silicate, DDSCAT coincides with the Mie solution in the whole
wavelength range that we studied ($0.03 - 100~\mu$m) while MarCoDES again
differs from the Mie solution in the region $\lambda \la 0.6~\mu$m.

 \begin{figure}
 \plotone{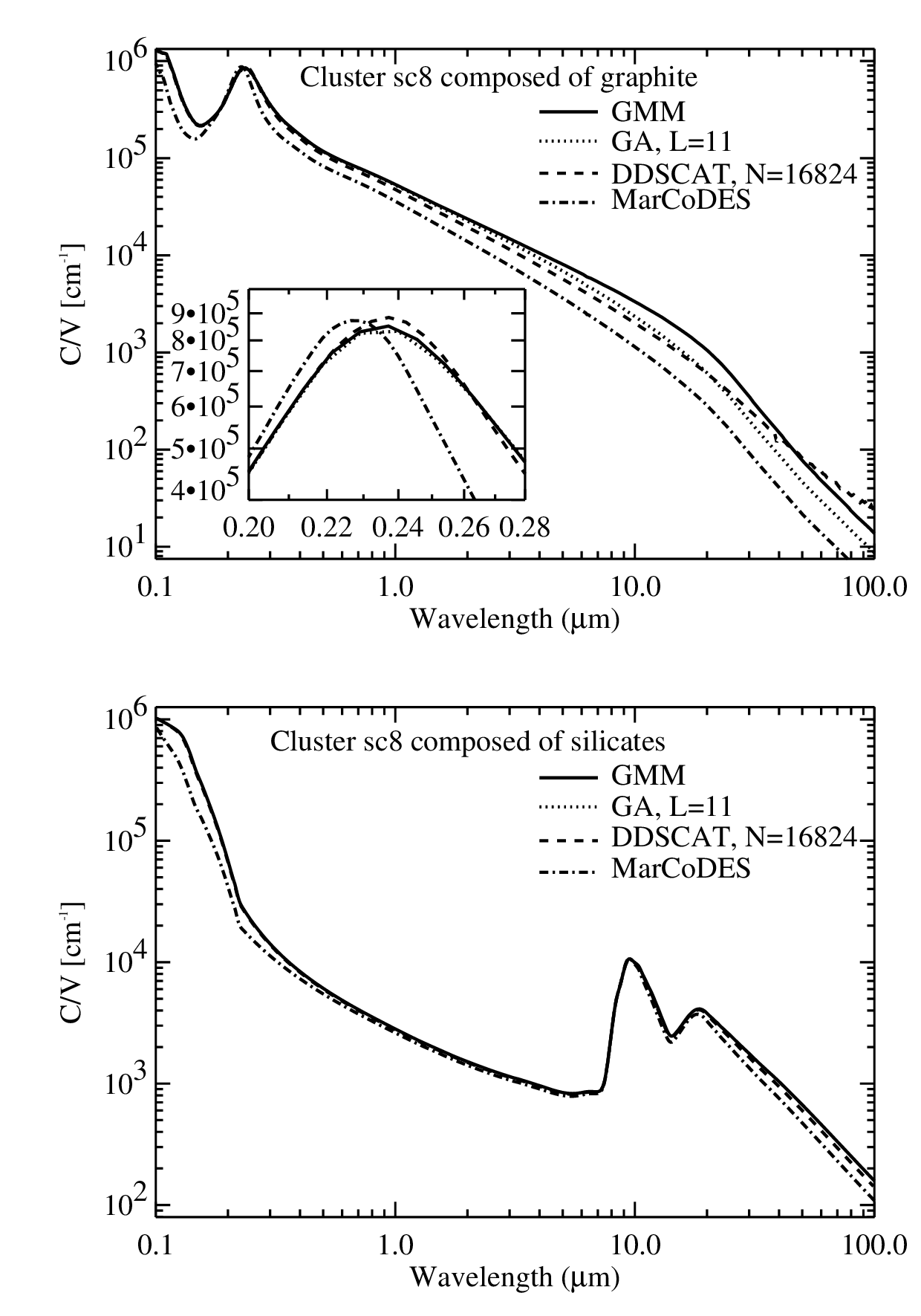}
 \caption{Cluster sc8 composed of graphitic material and silicates.
 The GMM solution has been computed with up to L=45 at certain 
 wavelengths; the GA, with up to L=11. The DDSCAT calculation was done with 
$32\times32\times32 = 32768$ dipoles (equivalent to 16824 dipoles within 
the target and 2103 dipoles per particle).}
 \end{figure}

Next we study the effect of clustering in the computation of the
extinction of graphitic and silicate clusters. As for single 
spheres, graphitic clusters allow us to better understand the nuances of
the different computational methods.
For the frac7 cluster (Fig.\,6) and the sc8 cluster (Fig.\,7) the GMM and GA
solution completely agree up to $\lambda \la 2.0~\mu$m showing that the
GA solution is indeed converged within the UV-vis region of the spectrum for
these clusters. At wavelengths $\lambda \ge 2.0~\mu$m the GA is not fully
converged as can be seen from the fact that it underestimates the extinction
for both clusters at longer wavelengths. DDSCAT agrees within 5\% with
the exact solution up to $\lambda \la 1.0~\mu$m at longer wavelengths it
underestimates the extinction but only slightly more than the not fully
converged GA solution. MarCoDES is 5\% off compared to GMM at $\lambda
= 0.24~\mu$m for the graphitic frac7 cluster and 10\% off for the 
graphitic sc8 cluster at the same wavelength. At wavelengths shortward 
and longward of $\lambda= 0.24~\mu$m MarCoDES significantly underestimates
the extinction. For the graphitic frac7 cluster (Fig.\,6) 
MarCoDES performs better
than for the sc8 cluster (Fig.\,7).  According to Markel et al.\ (2000), 
the performance of MarCoDES can be improved by altering
the intersection parameter which determines if the particles touch or
overlap, but we have not investigated that here. Nevertheless, the fact that 
MarCoDES gives higher extinction than GMM in some cases and lower in others,
suggests that the determination of
the rather arbitrary optimal intersection parameter is a very complex 
problem indeed.

For the silicate frac7 cluster (Fig.\,7) 
both the GA, DDSCAT and MarCoDES coincides completely
with the exact solution for $\lambda \la 11~\mu$m.  At longer wavelengths the
GA and DDSCAT solution coincide but overestimate the extinction by 
about 5\% while MarCoDES understimates the extinction at long wavelenghts for 
the frac7 cluster. For the sc8 silicate cluster (Fig.\,8) 
DDSCAT and GA coincide with the exact solution while MarCoDES 
underestimates the extinction at
$\lambda \la 5~\mu$m and $\lambda \ga 10~\mu$m. This suggests that the GA solution is fully converged for the silicate frac7 cluster 
within the whole wavelength
region considered and for the silicate sc8 cluster for $\lambda \la 11~\mu$m.
For the silicate clusters the same number of dipoles were used as for the
graphitic clusters in the DDSCAT calculations. 
DDSCAT deviates less than 5\% from the exact solution 
for both of the silicate clusters showing that it performs much better 
for materials with smaller refractive indices's than graphite.

\subsubsection{Computation time}

All of them, GA, GMM and DDSCAT required fairly large amounts of computer 
time; we point out, however, that all calculations were done on single 
processor machines (typically
with 800 MHz and 256 MB memory) and took at most a few days. 
The computation time is determined by the accuracy required, 
and even for a reasonable accuracy it is necessary to use a very fine 
discretization - i.e.\ 
a lot of dipoles for the DDA and high multi-pole orders for the GA and GMM.
This leads to a large number of linear equations which needs to be solved for 
the determination of the scattered electromagnetic field since the scattering
matrix is obtained by averaging the scattering matrices over a large number
of individual particles. Generally the computation time for 
a graphitic cluster was 3 times higher than that for the equivalent 
silicate 
cluster because of the much higher refractive index of graphite. MarCoDES 
is by far the fastest of all the methods but its accuracy is sometimes low, especially for compact clusters.

Regarding documentation, DDSCAT, MarCoDES and GMM have well documented user 
guides which make these programs fairly user friendly. For the GMM code, 
however, we needed some 
clarifying correspondence with its author. A new version of 
DDSCAT is now released (DDSCAT 6.0) which among other things 
have MPI capability for parallel computations of different target 
orientations (at a single wavelength); this should be very useful for 
calculations of averages over orientations (B.T.\,Draine pers.\,com.).
GMM and MarCoDES are also continuously being improved by their authors.

\section{Compact clusters vs. prefractal clusters}

 \begin{figure}
 \plotone{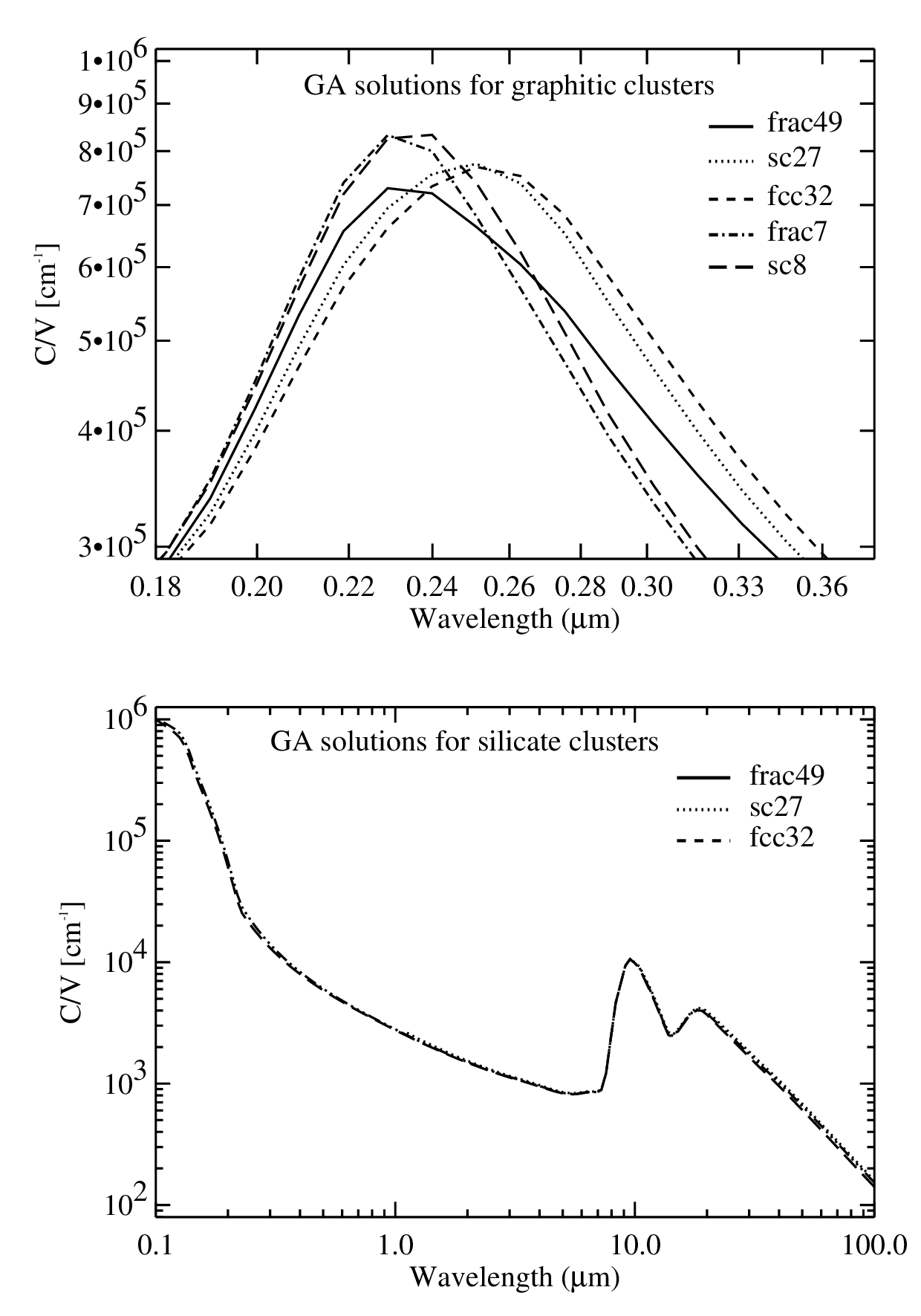}
 \caption{Extinction of prefractal and compact clusters calculated with the 
GA approach. The polar orders of the calculations were L=11 for the frac7 
and sc8 cluster and L=6 for the frac49, fcc32 and sc27 clusters. The top
figure shows the peaks of the graphitic clusters around 2200~{\AA}; it shows,
in particular, that the peaks of the prefractal clusters frac7 and 
frac49 are located quite close to one another.
The lower figure shows the extinction of silicate clusters. Regardless of
the material, the extinction of the different 
clusters are all of the same order of magnitude.} 
 \end{figure}

 \begin{figure}
 \plotone{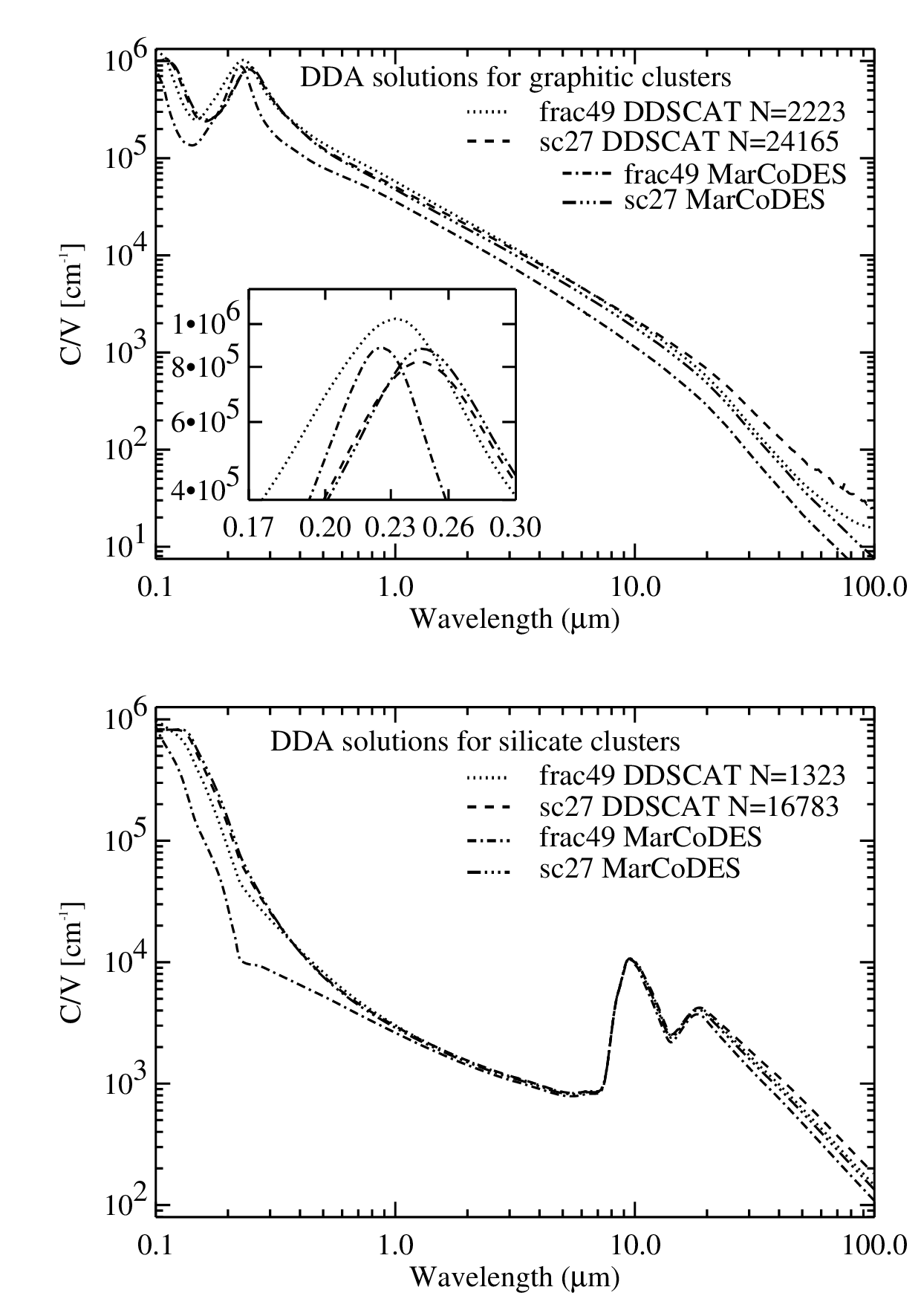}
 \caption{The extinction of prefractal and compact clusters as calculated by 
the DDA methods. 
Shown are the calculations for sc27 and frac49 performed with DDSCAT using
$32 \times 32 \times 32$ = 16783 dipoles and $36 \times 36 \times 36$ = 
46656 dipoles, which gave 32783 ($\sim 1214$ per particle) and 1323 ($\sim 331$
par particle) dipoles in the targets, respectively.
The figures show graphitic clusters at the top and silicate clusters at the 
bottom. These DDA calculations suggest that the extinction of
the different clusters are significantly more different than can be confirmed
by the GA calculations (see Fig.\,8).}
 \end{figure}

We now compare the extinction calculated with the different methods for prefractals and compact clusters.  

In Fig.\,8, the 
GA calculation for frac49 is compared to that of fcc32 and sc27 
around the 2200~{\AA} absorption feature. Here the conclusion would be (1) 
the prefractal cluster has a shift in
peak position and (2) the extinction of the prefractal clusters are of the same order of magnitude as the compact clusters. 
At long wavelengths all the considered cluster display an extinction of the same order of magnitude. 

Fig.\,9 shows the DDA calculations for the compact sc27 
cluster and the sparse frac49 cluster. A shift in peak position between the 
prefractal
and the compact cluster is observed around the 2200~{\AA} peak. DDSCAT
tends to indicate that the prefractal clusters have a somewhat enhanced extinction
around the 2200~{\AA} peak. At long wavelengths both codes show slightly
higher extinction for the compact sc27 cluster than for the frac49 cluster.

For the silicate clusters MarCoDES would lead to the conclusion that the 
prefractal clusters have
lower extinction at shorter wavelengths ($\lambda < 1~\mu$m) than the considered compact clusters while with
DDSCAT one would conclude that the extinction was of the same order of 
magnitude for the whole wavelength range in accordance with the GA result.

The GA calculations therefore suggest that the extinction of prefractal and
small compact clusters are on the same order of magnitude making it difficult 
to distinguish the different cluster morphology by observations. With the 
two DDA codes one might just as well reach the opposite (erroneous) conclusion. 

\section{Conclusions}

We have performed extinction calculations for clusters consisting of 
polycrystalline 
graphitic and silicate spheres in the wavelength range $0.1$ to $100~\mu$m.  
For the computations we have used the
rigorous multi-polar theory of G\'erardy \& Ausloos (1982; GA), the rigorous 
generalized
multi-particle Mie-solution by Xu (1995; GMM); the discrete dipole 
approximation using
one dipole per particle by Markel (1998; MarCoDES) and the discrete dipole 
approximation using multi dipoles by Draine \& Flatau (2000; DDSCAT). 

We have compared the extinction of open prefractal clusters and compact clusters.
The prefractal and small compact clusters display an extinction
of the same order of magnitude as when computed with the exact methods (GA and 
GMM). 
At shorter wavelengths around the 2200~{\AA} feature the graphitic 
prefractal clusters seem to have a stable peak position.

Overall, DDSCAT performs better than MarCoDES for all of the clusters.  
With DDSCAT, however, there is the 
unresolved question of how-many-dipoles are needed to
ensure a fairly accurate result, this number seems to follow a 
non-linear pattern so a more accurate result cannot always be
expected by doubling the number of dipoles (see Fig.\,5). 
MarCoDES is computationally much faster than the 
DDSCAT, GMM or GA method.  
The GMM computations were sufficiently fast so that convergence was reached over the whole studied wavelength range. On the other hand, 
our available GA program was slower and we could obtain converged results only in the UV-visible wavelength range.
Which of the four approaches is best to use for calculating the extinction 
of cluster particles will depend on the type of problem one wants to address 
and the accuracy needed.  

\acknowledgements 
We would like to thank B.T.\,Draine, V.A.\,Markel and Y-L.\,Xu for making their
codes available as shareware. J.S. acknowledges very helpful correspondence 
with Y-L.\,Xu regarding the use of GMM.

\end{document}